\documentclass[prb,twocolumn,showpacs,preprintnumbers,amsmath,amssymb]{revtex4}
\usepackage{graphicx}
\usepackage{dcolumn}
\usepackage{bm}

\begin{document}

\title{First--principles calculation of the intersublattice  exchange
 interactions and  Curie  temperatures of
full Heusler alloys Ni$_2$MnX (X$=$Ga, In, Sn, Sb)}

\author{E.~\c Sa\c s\i o\~glu, L. M.  Sandratskii and  P. Bruno}

\affiliation{Max-Planck Institut f\"ur Mikrostrukturphysik,
D-06120 Halle, Germany }

\date{\today}

\begin{abstract}

The interatomic exchange interactions and Curie temperatures in
Ni--based full Heusler alloys Ni$_2$MnX with X$=$Ga, In, Sn and Sb
are studied within the framework of the density-functional theory.
The calculation of the exchange parameters is based on the
frozen--magnon approach. Despite closeness of the experimental
Curie temperatures for all four systems their magnetism appeared
to differ strongly. This difference involves both the Mn--Mn and
Mn--Ni exchange interactions. The Curie temperatures, T$_C$, are
calculated within the mean--field approximation by solving a
matrix equation for a multi--sublattice system. Good agreement
with experiment for all four systems is obtained. The role of
different exchange interactions in the formation of T$_C$ of the
systems is discussed.

\end{abstract}


\maketitle

\section{introduction}
Much efforts is currently devoted to the search for ferromagnetic
materials suitable for application in semiconductor spintronics
devices. To allow an efficient spin-injection into semiconductor
these materials must satisfy a number of conditions. In particular
they must have the Curie temperature noticeably higher than the
room temperature, be compatible with the semiconductors used
industrially, and possess a very high spin-polarization of the
electron states at the Fermi level. \cite{g_ref1} Some of the
Heusler compounds were found to have half-metallic ground state
\cite{half_m1, half_m2} which is characterized by a 100\% carrier
spin-polarization.

A feature of other Heusler alloys that is useful for spintronic
applications is very small lattice mismatch with widely employed
semiconductors (e.g., Ni$_2$MnIn and InAs). \cite{inas_1, inas_2,
inas_3} Among further prominent physical properties of this class
of materials one can mention a martensitic transformation in
Ni$_2$MnGa which takes place  below the Curie temperature.
\cite{martensitic} An interesting combination of physical
properties makes Heusler alloys the subject of intensive
experimental and theoretical investigations which  could motivate
their use  as multifunctional materials in practical applications.
\cite{ref_5, ref_6, ref_7, ref_8, ref_9, ref_10, ref_11}

In the present work we report the theoretical study of the
exchange interactions and Curie temperature of four full Heusler
compounds Ni$_2$MnX with X$=$Ga, In, Sn and  Sb. Experimentally,
all of them are ferromagnetic and have similar values of the Curie
temperature.\cite{lattice}

First important contribution to the density  functional theory
study of these systems was made in early paper by K\"ubler
{\textit et al.}, \cite{Kubler2} where the microscopic mechanisms
of the magnetism of Heusler alloys was discussed on the basis of
the calculation of the ferromagnetic and antiferromagnetic
alignments of the Mn moments. Recently a detailed study of the
magnetic interactions in Ni$_2$MnGa and Ni$_2$MnAl  was reported
by Enkovaara {\textit  et al} \cite{enkovaara}. The authors used
the plane spiral configurations and have shown that the Ni
sublattice plays important role in the magnetic properties of the
system.

The main purpose of the present work is a detailed study of the
exchange interactions in these systems. In particular we report a
systematic study of the exchange interaction between atoms of
different sublattices and show that pattern of exchange
interactions in these systems deviates strongly from the physical
picture that can be expected on the basis of the experimental
information available. Indeed common crystal structure, similar
chemical composition and close experimental values of the Curie
temperature make the assumption natural that the exchange
interactions in these systems are similar. Our study shows,
however, that this assumption is not correct. The exchange
interactions vary strongly depending on the X constituent. In
particular, the inter-sublattice interactions change strongly from
system to system. We show that different exchange interactions
lead, in agreement with experiment, to similar values of the Curie
temperatures. We analyze the relation between the properties of
the exchange interactions and the Curie temperatures.

The  paper is  organized as follows. In Sec. II  we present the calculational
approach. Section III contains the results of the calculations and discussion.
Section IV gives the conclusions.

\section{calculational approach}

The calculations are carried out with the augmented spherical
waves  method \cite{asw} within the atomic--sphere
approximation.\cite{asa} The exchange--correlation potential
is chosen in the generalized gradient approximation.\cite{gga} A dense
Brillouin zone (BZ) sampling $30\times30\times30$ is used.
In all calculations the experimental values
of the lattice parameters are used (Table \ref{tab:moments}).
\cite{lattice} The radii of all atomic spheres are chosen equal.

We describe the interatomic exchange interactions in terms of the
classical Heisenberg Hamiltonian
\begin{equation}
\label{eq:hamiltonian2}
 H_{eff}=-  \sum_{\mu,\nu}\sum_{\begin {array}{c}
^{{\bf R},{\bf R'}}\\ ^{(\mu{\bf R} \ne \nu{\bf R'})}\\
\end{array}} J_{{\bf R}{\bf R'}}^{\mu\nu}
{\bf s}_{\bf R}^{\mu}{\bf s}_{\bf R'}^{\nu}
\end{equation}
In Eq.(\ref{eq:hamiltonian2}), the  indices  $\mu$ and $\nu$
number different sublattices and ${\bf R}$ and ${\bf R'}$ are the
lattice vectors specifying the atoms within sublattices, ${\bf
s}_{\bf R}^\mu$ is the unit vector pointing in the direction of
the magnetic moment at site $(\mu,{\bf R})$. The systems
considered contain three 3d atoms in the unit cell with positions
$(\frac{1}{4}, \frac{1}{4},\frac{1}{4})$ for the Mn atom and
$(0,0,0)$ and $(\frac{1}{2},\frac{1}{2},\frac{1}{2})$ for two Ni
atoms.

We employ the frozen--magnon approach \cite{magnon_1, magnon_2,
magnon_3} to calculate interatomic Heisenberg exchange parameters.
The calculations involve few steps. In the first step, the
exchange parameters between the atoms of a given sublattice $\mu$
are computed. The calculation is based on the evaluation of the
energy of the frozen--magnon configurations defined by the
following atomic polar and azimuthal angles
\begin{equation}
\theta_{\bf R}^{\mu}=\theta, \:\: \phi_{\bf R}^{\mu}={\bf q \cdot
R}+\phi^{\mu}. \label{eq_magnon}
\end{equation}
In the calculation discussed in this paper the constant phase
$\phi^{\mu}$ is always chosen equal to zero. The magnetic moments
of all other sublattices are kept parallel to the z axis. Within
the Heisenberg model~(\ref{eq:hamiltonian2}) the energy of such
configuration takes the form
\begin{equation}
\label{eq:e_of_q} E^{\mu\mu}(\theta,{\bf
q})=E_0^{\mu\mu}(\theta)+\sin^{2}\theta J^{\mu\mu}({\bf q})
\end{equation}
where $E_0^{\mu\mu}$ does not depend on {\bf q} and the Fourier transform $J^{\mu\nu}({\bf q})$
is defined by
\begin{equation}
\label{eq:J_q}
J^{\mu\nu}({\bf q})=\sum_{\bf R}
J_{0{\bf R}}^{\mu\nu}\:\exp(i{\bf q\cdot R}).
\end{equation}
In the case of $\nu=\mu$ the sum in Eq. (\ref{eq:J_q}) does not
include ${\bf R}=0$. Calculating $ E^{\mu\mu}(\theta,{\bf q})$ for
a regular ${\bf q}$--mesh in the Brillouin zone of the crystal and
performing back Fourier transformation one gets exchange
parameters $J_{0{\bf R}}^{\mu\mu}$ for sublattice $\mu$.

To determine the exchange interactions between the atoms of two
different sublattices $\mu$ and $\nu$ the frozen--magnon
configurations (Eq. \ref{eq_magnon}) are formed for both
sublattices. The Heisenberg energy of such configurations takes
the form
\begin{eqnarray}
E^{\mu\nu}(\theta,{\bf q})&=&E_0^{\mu\nu}(\theta)+\sin^{2}{\theta}
\:[J^{\mu\mu}({\bf q})+J^{\nu\nu}({\bf q})]\nonumber \\
&& +2\sin^{2}{\theta}\:{\text Re} J^{\mu\nu}({\bf q})
\label{eq:sublatticenergy}
\end{eqnarray}
where $E_0^{\mu\nu}(\theta)$ is a ${\bf q}$--independent part.
Performing calculation of $ [E^{\mu\nu}(\theta,{\bf q})-
E^{\mu\nu}(\theta,{\bf 0})]$ and subtracting single-sublattice
contributions known from the previous step one finds $[\:{\text
Re} J^{\mu\nu}({\bf q})-{\text Re} J^{\mu\nu}({\bf 0})]$. The back
Fourier transformation of this expression gives for ${\bf R}\ne0$
the following combinations of the interatomic exchange parameters:
$J_{\bf R}^{\mu\nu}\equiv\frac{1}{2}(J_{0{\bf
R}}^{\mu\nu}+J_{0{\bf (-R)}}^{\mu\nu})$. In general, one needs to
perform similar calculations for different phases $\phi^{\mu}$ in
Eq.~(\ref{eq_magnon}) to get another linear combination of the
parameters to be able to separate them. For the systems considered
these calculations can be avoided by taking into account the
symmetry of the lattice (see below).

Quantities $J_{\bf R}^{\mu\nu}$ does not contain information about
the interaction of the atoms  within the first unit cell
corresponding to  ${\bf R}=0$. These exchange parameters can be
found in the calculations for magnetic configurations periodic
with the periodicity of the lattice. The atoms in the unit cell
are separated into two groups. Within each group the moments are
parallel. The moments from the different groups form an angle
$\theta$. The energies of such magnetic configurations are
expressed through the sums $J_0^{\mu\nu}\equiv\sum_{\bf R}
J_{0{\bf R}}^{\mu\nu}$. Since the sums $\sum_{\bf R \ne 0}
J_{0{\bf R}}^{\mu\nu}$ are known from the preceding step the
parameters with ${\bf R}=0$ become accessible. The symmetry
relation $J_{00}^{\mu\nu}= J_{0{\bf R}}^{\mu\nu}$ for ${\bf R}=(0,
\frac{1}{2}, \frac{1}{2})$ allows to split the sums $J_{{\bf
R}}^{\mu\nu}$ to individual parameters.
\begin{figure}[t]
\begin{center}
\includegraphics[scale=0.5]{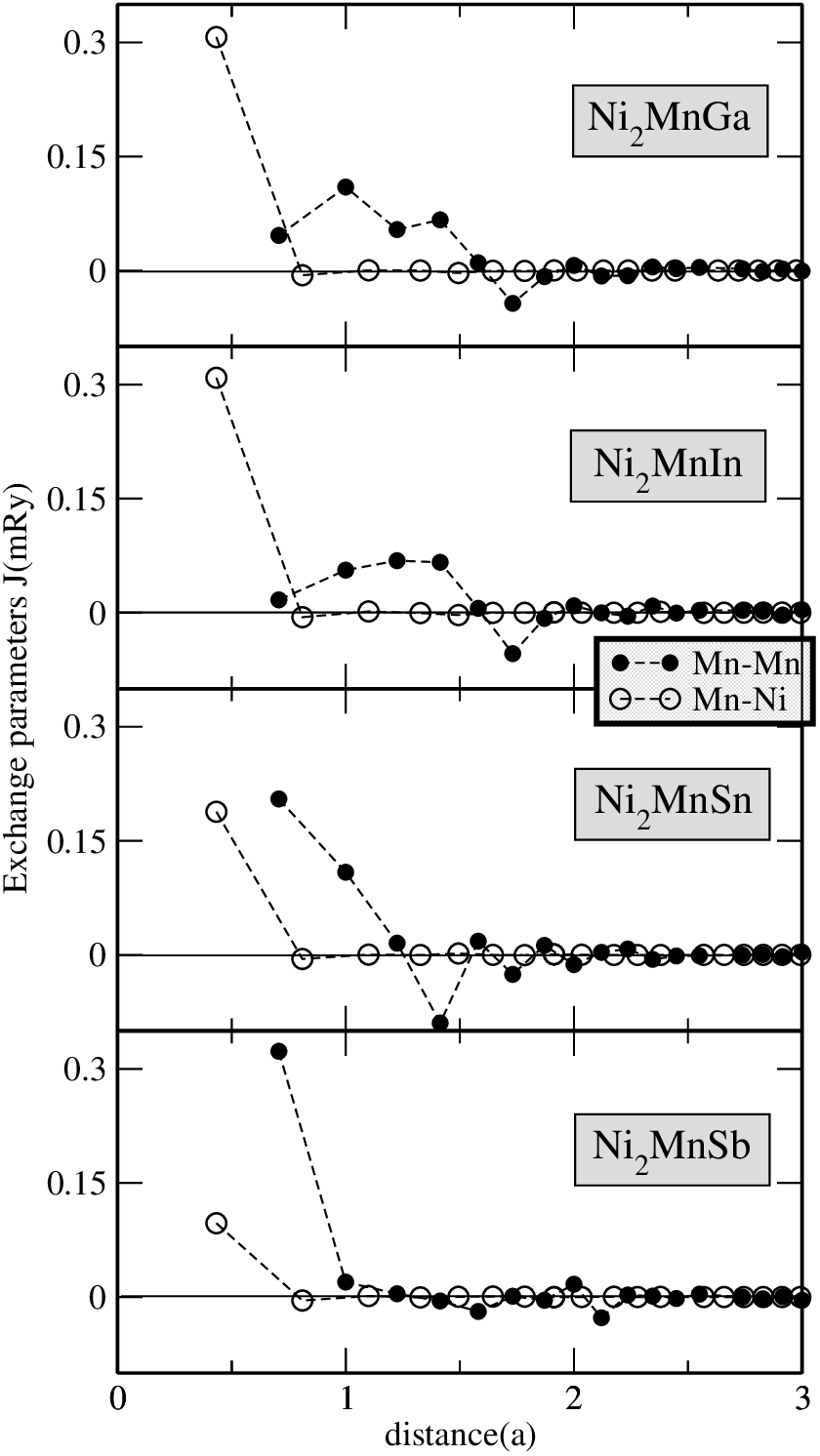}
\end{center}
\caption{The parameters of the Mn--Mn (intra--sublattice) exchange interactions
and  Mn--Ni (inter--sublattice) exchange interactions in  Ni$_2$MnX
(X$=$Ga, In, Sn, Sb). The distances
are given in the units of the lattice  constant.
The significance of the oscillations of the exchange parameters
is verified by varying the {\bf q} mesh in the frozen-magnon calculations.}
\label{exchange}
\end{figure}

The Curie  temperature is estimated  within the mean--field
approximation for a multi--sublattice material by solving the
system of  coupled   equations \cite{Anderson}
\begin{equation}
\label{eq_system}
\langle s^{\mu}\rangle
=\frac{2}{3k_BT}\sum_{\nu}J_0^{\mu\nu}\langle s^{\nu}\rangle
\end{equation}
where  $\langle s^{\nu}\rangle$ is the average $z$ component of
${\bf s}_{{\bf R}}^{\nu}$. Eq. (\ref{eq_system}) can be
represented in the form of eigenvalue matrix problem
\begin{equation}
\label{eq_eigenvalue} ({\bf \Theta}-T {\bf I}){\bf S}=0
\end{equation}
where $\Theta_{\mu\nu}=\frac{2}{3k_B}J_0^{\mu\nu}$, ${\bf I}$ is a
unit matrix and ${\bf S}$ is the vector of $\langle s^{\nu}\rangle
$. The largest eigenvalue of matrix $\Theta$ gives the value of
Curie temperature. \cite{Anderson}

\section{results and  discussion}

In Table \ref{tab:moments} we present calculated magnetic moments.
For comparison, the available experimental values of the moments and the
results of previous calculations are presented. Relative variation of the Mn moment is small.
On the other hand, the moment of Ni and X
atoms show strong relative variation and are in Ni$_2$MnSb about
two times smaller than in Ni$_2$MnGa or Ni$_2$MnIn. The values of the
magnetic moments are in  good agrement with the results of previous
calculations.

\begin{table}
\caption{Experimental lattice parameters  and magnetic moments (in
 $\mu_B$) of Ni$_2$MnX (X$=$Ga, In, Sn, Sb). }
\begin{ruledtabular}
\begin{tabular}{lccccc}
$$&a(a.u.)& Mn & Ni &  X & Cell
\\ \hline \\
Ni$_2$MnGa  &11.058$^a$& 3.570  & 0.294 & -0.068 & 4.090 \\
            && (3.43$^b$)   & (0.36$^b$)  &(-0.04$^b$)  & (4.11$^b$)\\
            &&    &   &   & (4.17$^c$) \\

Ni$_2$MnIn &11.468$^a$ & 3.719  & 0.277 & -0.066 & 4.208 \\

Ni$_2$MnSn &11.419$^a$ & 3.724  & 0.206 & -0.057 & 4.080 \\
            && (3.53$^d$)   & (0.24$^d$)  &  (-0.03$^d$) & (4.08$^d$)\\

Ni$_2$MnSb &11.345$^a$ & 3.696  & 0.143 & -0.033 & 3.950 \\
\end{tabular}
\end{ruledtabular}

$^a$Ref.\onlinecite{lattice}\\
$^b$Ref.\onlinecite{magmo1}\\
$^c$Ref.\onlinecite{magmo2} (Exp.)\\
$^d$Ref.\onlinecite{magmo3}\\

 \label{tab:moments}
\end{table}

\begin{figure}[t]
\begin{center}
\includegraphics[scale=0.36]{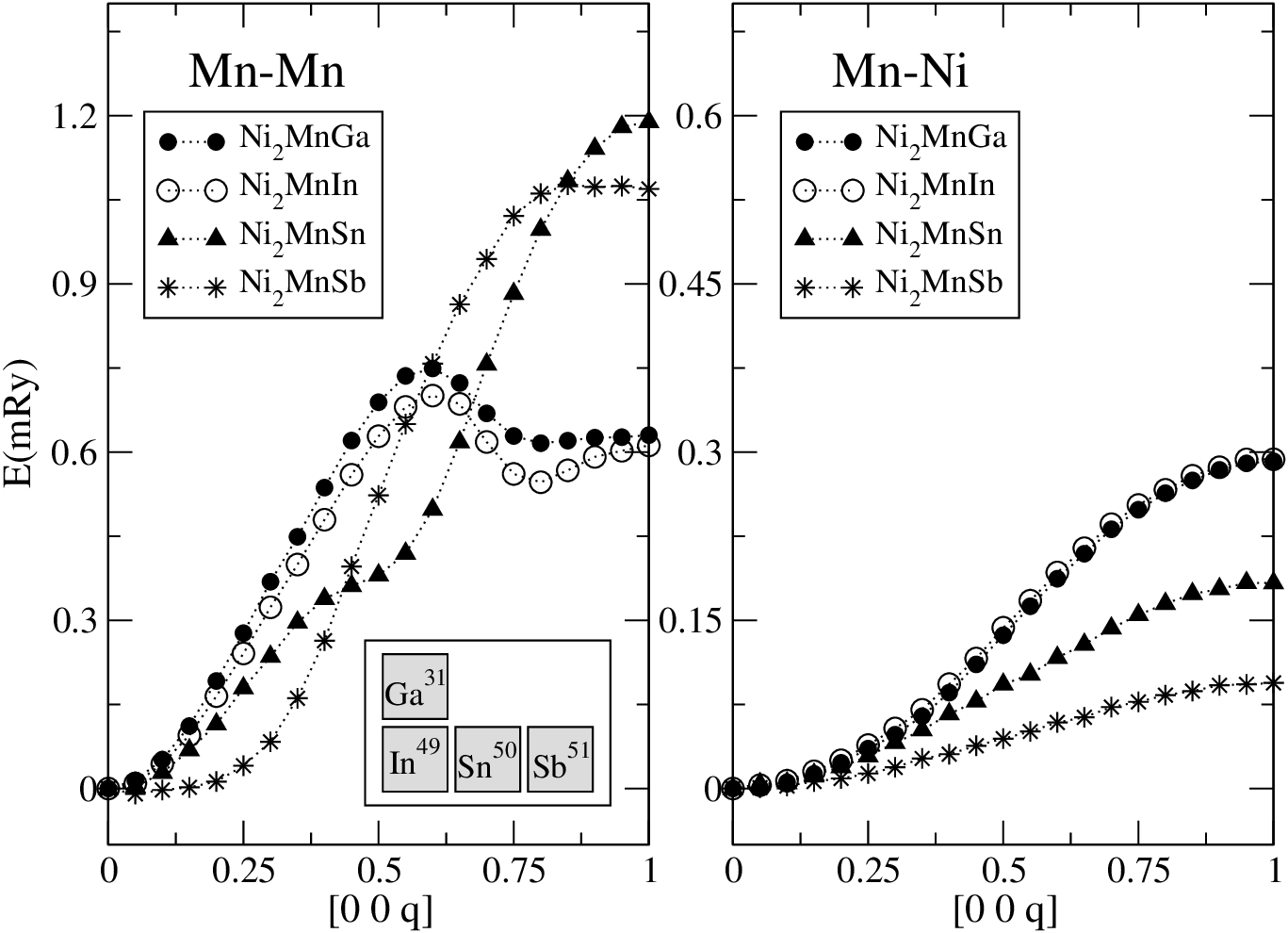}
\end{center}
\caption{Frozen--magnon  energies  as a  function  of the wave
vector ${\bf q}$ (in units of $2\pi/a$)  in Ni$_2$MnX for Mn--Mn
(left) and Mn-Ni interactions (right).
} \label{total_energies}
\end{figure}

The  calculated Heisenberg exchange  parameters are presented  in
figure \ref{exchange}. As mentioned in the introduction, the
assumption that the closeness of the experimental Curie
temperatures is the consequence of the similarity of the exchange
interactions is not confirmed by the calculations. We obtain
strong dependence of the exchange interactions on the type of the
X atom. For X=Ga and X=In that belong to the same column of the
Mendellev's table (see inset in Fig. \ref{total_energies}) we
obtain similar pattern of Heisenberg exchange parameters. On the
other hand, for X atoms belonging to different columns the changes
in the exchange interactions are very strong (Fig. \ref{exchange}).
These changes concern both the Mn--Mn intra-sublattice
interactions and the Ni-Mn inter-sublattice interaction.

Considering the Mn--Mn interactions we notice that in Ni$_2$MnGa
and Ni$_2$MnIn the interaction with the coordination spheres from
the first to the forth is positive. The interaction with the first
coordination sphere is weaker than with the following ones.  The
interaction with the fifth sphere is very small. The interaction
with the 6th sphere is negative. The interaction with further
coordination spheres is very weak.

In Ni$_2$MnSn the interaction with the first sphere strongly
increases compared with Ni$_2$MnGa and Ni$_2$MnIn. On the other
hand, the interaction with the third sphere becomes small. The
interaction with the forth sphere is strongly negative. The
interaction with further neighbors are weak.

The trend observed in transition from Ni$_2$MnGa and Ni$_2$MnIn to
 Ni$_2$MnSn becomes even stronger in the case of X=Sb. Here the
interaction with the first neighbor increases further and is the
only strong exchange interaction between the Mn atoms.

The inter-sublattice Mn--Ni interaction behaves very differently.
A sizable interaction takes place only between nearest neighbors.
This interaction is very strong in  Ni$_2$MnGa and Ni$_2$MnIn and
quickly  decreases for X=Sn and, especially, X=Sb.

\begin{figure}[t]

\begin{center}
\includegraphics[scale=0.32]{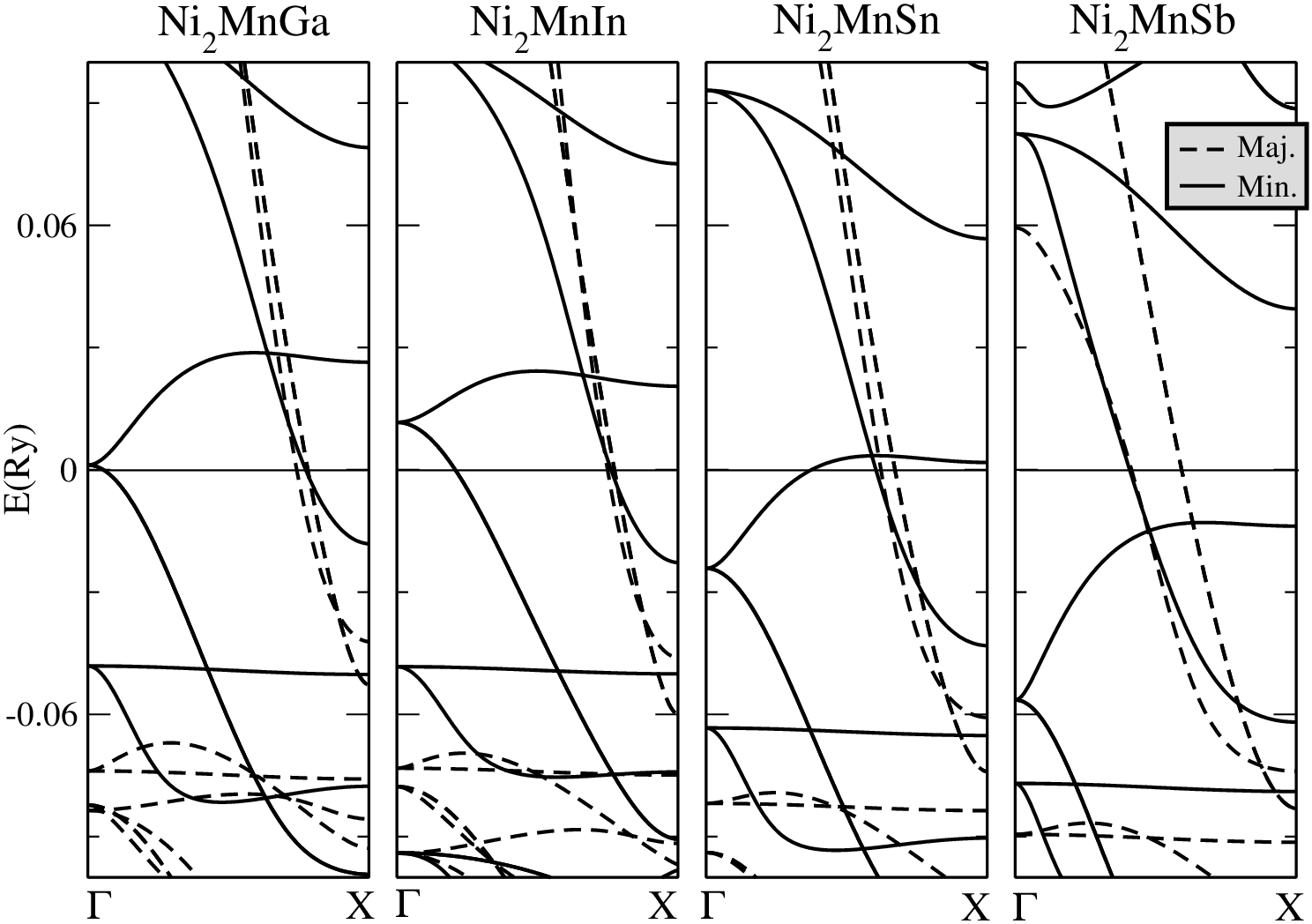}
\end{center}
\caption{Band structure  of Ni$_2$MnX along  the  high symmetry
line ($\Gamma$--X).
} \label{band_structure}
\end{figure}
\begin{figure}[t]
\begin{center}
\includegraphics[scale=0.43]{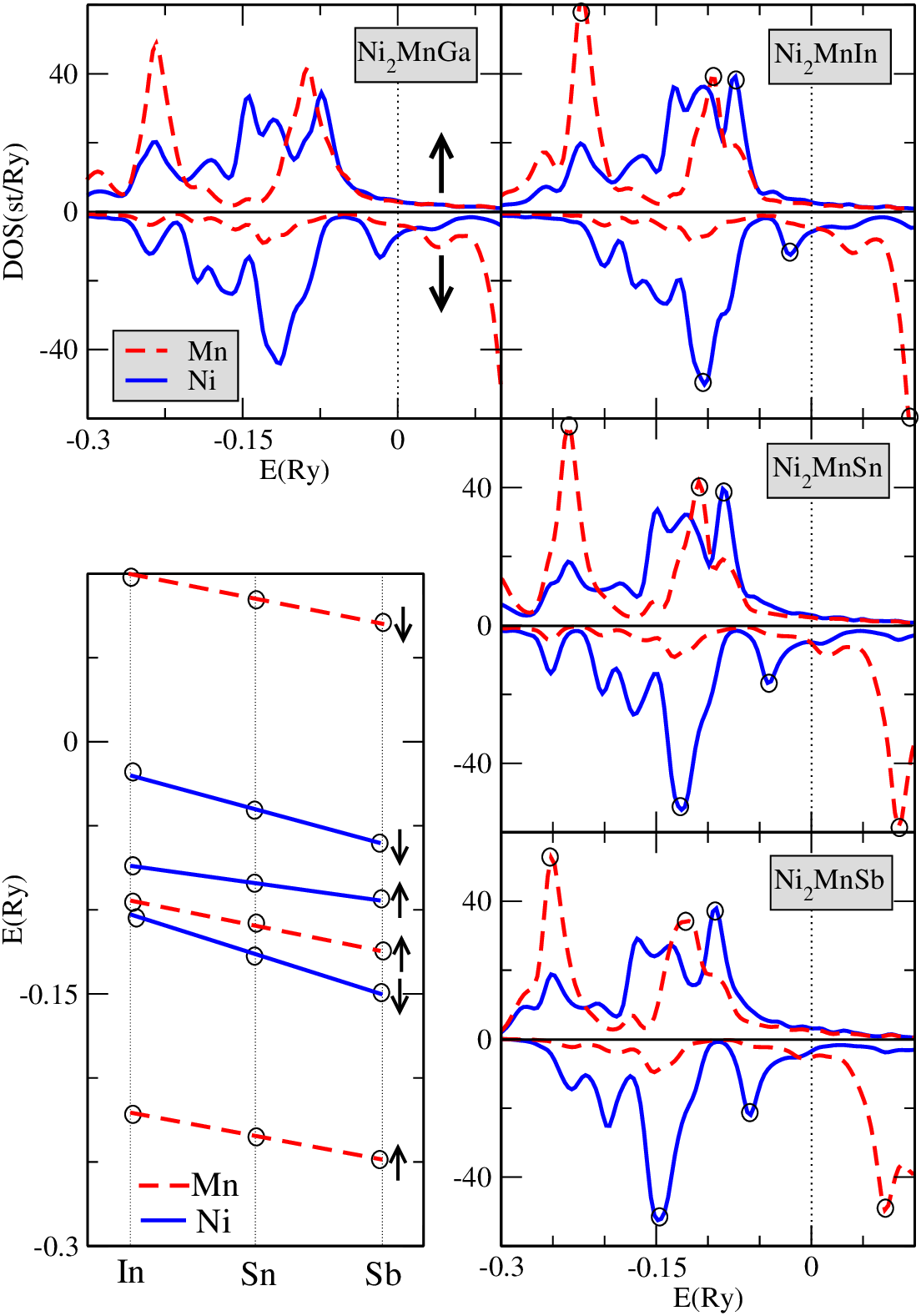}
\end{center}
\caption{ Spin--projected partial density of states  of Ni$_2$MnX.
The separate graph in the left-hand bottom part of the figure shows
the variation of the positions of a number of the DOS peaks
for the In-Sn-Sb series of compounds. The lines in this graph are
guides for the eye. The energies
are measured with respect to the Fermi level of the corresponding
system. The small open circles on the DOS curves mark the positions
of the peaks. Arrows show spin projections.
The analysis of the states at the Fermi level shows that the
main contributions comes from the 3d states of Mn and Ni.
}
\label{partial_dos}
\end{figure}

To reveal the physical origin of the strong difference in the
exchange parameters of these systems we plot in Fig.
\ref{total_energies} the frozen-magnon energies as a function of
wave vector {\bf q} for one direction in the Brillouin zone.\cite{magnonenergies} We remind
the reader that the inter-atomic exchange parameters are Fourier
transforms of the E(${\bf q}$) functions and therefore reflect
their form. Indeed,  E(${\bf q}$) curves for Ni$_2$MnGa and
Ni$_2$MnIn are close to each other that leads to a similar set of
interatomic exchange parameters (Fig. \ref{exchange}). These
curves deviate strongly from a simple cosinusoid having a maximum
at ${\bf q}$ about $0.6$ and a rather weak variation at ${\bf
q}>0.6$. \cite{phonon} The complexity of E(${\bf q}$)
means that several Fourier components are needed to describe the
features of the function. This is reflected in the the
Heisenberg's parameters of Ni$_2$MnGa and Ni$_2$MnIn.

On the other hand, the E(${\bf q}$) curve of  Ni$_2$MnSb is well
described by one cosinusoid (Fig. \ref{total_energies}) that
results in a single large Mn--Mn exchange parameter
(Fig.~\ref{exchange}). The E(${\bf q}$) of Ni$_2$MnSn assumes an
intermediate position from the viewpoint of the complexity of the
function. This property is also reflected in the exchange parameters
(Fig.~\ref{exchange}).

Note that the character of the ${\bf q}$--dependence of the energy
illustrated by Fig.~\ref{total_energies} is a consequence of the
properties of the electronic structure of the compounds. Indeed,
in Fig.~\ref{band_structure} we see that the electronic structures
of Ni$_2$MnGa and Ni$_2$MnIn are similar. Transition along the row
In--Sn--Sb leads to increasing difference in the electron
spectrum. This increasing difference can be traced back to the
change in the number of valence electrons: a Sb atoms has two more
valence electrons than In and one more electron than Sn. As the
result an important difference in the electron structure of the
system is a relative shift of the Fermi level to a higher energy
position in the sequence In--Sn--Sb.  This shift is clearly seen
in the DOS presented in  Fig.~\ref{partial_dos}. The positions of
the same features of the DOS in different systems are well
described by linear functions with negative angle coefficients.
For the Mn peaks all three lines are almost parallel. This means
that the change in the Mn DOS from system to system can be treated
as a rigid shift with respect to the Fermi level. In the case of
the Ni-DOS the situation is more complicated since, besides the
variation of the electron number, an additional influence on the
peak positions is exerted by the variation of the Ni atomic moment
(see Table \ref{tab:moments}).

The E(${\bf q}$) curves determining the Mn--Ni interactions are
presented in Fig. \ref{total_energies}. The form of the curves is
in all cases close to a cosinusoid. Therefore, only one exchange
parameter has sizable value. The strength of the interaction is in
correlation with the value of the magnetic moment of the Ni atoms.

\begin{table}
\caption{Mean--field estimation of the Curie temperatures  for
Ni$_2$MnX (X$=$Ga, In, Sn, Sb). The experimantal Curie temperature
values are taken  from Ref.\onlinecite{lattice}.
\label{tab:Curietemperature}}
\begin{ruledtabular}
\begin{tabular}{lccc}
& $T_{c}^{{Mn-Mn}[MFA]}(K)$ & $T_{c}^{[MFA]}(K)$ & $ T_{c}^{[Exp]}(K)$ \\ \hline \\
Ni$_2$MnGa  & 302   & 389  & 380  \\
Ni$_2$MnIn  & 244   & 343  & 315  \\
Ni$_2$MnSn  & 323   & 358  & 360  \\
Ni$_2$MnSb  & 343   & 352  & 365  \\
\end{tabular}
\end{ruledtabular}
\end{table}

The interatomic exchange parameters are used to evaluate the Curie
temperature. If only the Mn-Mn interactions are taken into account
we obtain values shown in Table \ref{tab:Curietemperature}.
Despite very strong difference in the Mn--Mn exchange parameters
in these systems the difference in the corresponding Curie
temperatures is not very large. The explanation for this result is
the property that in MFA to a one--sublattice ferromagnet the
value of the Curie temperature is determined by the sum of the
interatomic exchange interactions $J_0=\sum_{{\bf R}\ne 0}
J_{0{\bf R}}$. $J_0$ gives the average value of E(${\bf q}$) and
is less sensitive to the detailed form of the E(${\bf q}$)
function.

The comparison of the Curie temperatures calculated with the use
of the Mn--Mn exchange parameters only with experimental Curie
temperatures shows that the agreement with experiment is not in
general good. In the case of Ni$_2$MnGa the error is about 30\%.

Account for inter--sublattice interactions improves the agreement with
experimental $T_C$ values
considerably (Table \ref{tab:Curietemperature}).
This shows that the Ni moment provides a magnetic
degree of freedom which plays important role in the thermodynamics
of the system.

\section{conclusion}

In conclusion, we  have  systematically  studied exchange
interactions and Curie temperatures in Ni--based   full  Heusler
alloys Ni$_2$MnX (X= Ga, In, Sn, Sb) within the  parameter--free
density functional theory. Our  calculations show that despite
similarity of the Curie temperatures of these systems there is
strong difference in the underlying magnetic interactions. This
difference involves both the Mn--Mn and Mn--Ni exchange
interactions which depend strongly on the X constituent. The Curie
temperatures are calculated within the mean--field approximation
to the classical Heisenberg Hamiltonian by solving a matrix
equation for a multi--sublattice system. Good agreement with
experiment for all four systems is obtained.

\begin{acknowledgments}
The financial support of Bundesministerium f\"ur Bildung und
Forschung is acknowledged.
\end{acknowledgments}


\begin{thebibliography}{99}
\bibitem{magnonenergies}
The energy of the frozen magnon with a given ${\bf q}$ can be seen
as a result of a complex interaction of the ferromagnetic states
separated by vector ${\bf q}$ in the reciprocal space [L. M.
Sandratskii, Adv. Phys. {\bf 47}, 1 (1998)]. This interaction is
stronger if the states have close energies and weaker for the
states separated by a large energy interval. Note that the DOS is
an integrated characteristic and does not contain information
about the distribution of the states in the reciprocal space.
Therefore the study of the DOS is not sufficient for understanding
of the character of the ${\bf q}$ dependence of the frozen-magnon
dispersion.
\end{thebibliography}
\end{document}